\documentclass[conference]{IEEEtran}

\usepackage[absolute,showboxes]{textpos}
\newcommand{\copyrightstatement}{
\begin{textblock}{0.84}(0.08,0.95)    
\vspace{2mm}
\centering This work has been accepted for presentation at IEEE LATINCOM 2020. 978-1-7281-8903-1/20/\$31.00~\copyright{}2020 IEEE.
\vspace{2mm}
\end{textblock}
}

\usepackage{cite}
\usepackage{amsmath,amssymb,amsfonts}
\usepackage{algorithmic}
\usepackage{graphicx}
\usepackage{textcomp}
\usepackage{xcolor}
\usepackage{comment}
\usepackage{mathtools}
\usepackage{xspace}
\usepackage{comment} 

\usepackage{tikz}

\newcommand{\sep}{\hspace{2 mm}}
\def\BibTeX{{\rm B\kern-.05em{\sc i\kern-.025em b}\kern-.08em
\kern-.1667em\lower.7ex\hbox{E}\kern-.125emX}}
    
\usepackage{comment} 
\usepackage[english,ruled,linesnumbered,vlined]{algorithm2e}
\usepackage{setspace}
\usepackage{caption}
\usepackage{array}
\usepackage{booktabs} 
\usepackage{bigstrut} 
\usepackage{boldline} 
\usepackage{multirow} 
\usepackage{threeparttable}
\usepackage[normalem]{ulem} 
\usepackage{anyfontsize}
\usepackage{siunitx} 

\definecolor{ao(english)}{rgb}{0.0, 0.5, 0.0}

\definecolor{color1}{RGB}{198, 90, 103}
\definecolor{color2}{RGB}{175, 42, 48}
\definecolor{color3}{RGB}{191, 6, 29}
\definecolor{color4}{RGB}{179, 43, 204} 	 
\definecolor{color5}{RGB}{0,0,0}           
\newcommand{\agnd}[1]{\textcolor{black}{#1}}
\newcommand{\ca}[1]{\textcolor{black}{#1}}
\newcommand{\agnc}[1]{\textcolor{black}{#1}}
\newcommand{\agnb}[1]{\textcolor{black}{#1}}
\newcommand{\agn}[1]{\textcolor{black}{#1}}
\newcommand{\cp}[1]{\textcolor{black}{#1}}

\title{
GROWN: Local Data Compression in Real-Time \\ To Support Energy Efficiency in WBAN
}

\author{
\IEEEauthorblockN{{\bf Cainã Passos\IEEEauthorrefmark{1}}, {\bf Carlos Pedroso\IEEEauthorrefmark{1}}, {\bf Agnaldo Batista\IEEEauthorrefmark{1}}, {\bf Michele Nogueira\IEEEauthorrefmark{1}, \bf Aldri Santos\IEEEauthorrefmark{1}}
}
\IEEEauthorblockA{\IEEEauthorrefmark{1}Wireless and Advanced Networks Laboratory (NR2) - UFPR, Brazil  \\
Emails:\{cpassos, capjunior, asbatista, michele, aldri\}@inf.ufpr.br 
}}

\IEEEaftertitletext{\vspace{-1.0\baselineskip}}

\begin{document}

\copyrightstatement

\maketitle


\begin{abstract}

The evolution of wireless technologies has enabled the creation of networks for several purposes as health care monitoring. The Wireless Body Area Networks (WBANs) enable continuous and real-time monitoring of physiological signals, but that monitoring leads to an excessive data transmission usage, and drastically affects the power consumption of the devices. Although there are approaches for reducing energy consumption, many of them do not consider information redundancy to reduce the power consumption. This paper proposes a hybrid approach of local data compression, called GROWN, to decrease information redundancy during data transmission and reduce the energy consumption. Our approach combines local data compression methods found in WSN. We have evaluated GROWN by experimentation, and the results show a decrease in energy consumption of the devices and an increase in network lifetime.

\end{abstract}

\section{Introduction}
\label{sec:int}
\cp{
Wireless sensors networks (WSN) and nanosystems have enabled the deployment of wearables sensors added into  clothing or implanted in the body, such as fitness trackers and smartwatches~\cite{tavares2020traffic}. Wireless body area networks (WBANs) enables the collect, monitoring, and transmission of physiological signals from people to several medical applications and health professionals, making patients day-to-day easier~\cite{movassaghi2014wireless}~\cite{resque2019assessing}.
The WBAN devices usually employ small~batteries, which~limit the energy consumption to communication~\cite{movassaghi2014wireless}. Further, 
wearable and implantable devices working on real-time should have a long lifetime, mainly due to challenges for battery recharge and/or replacement~\cite{liao2018mutual}. Thus, manners for saving transmission energy can support 
the services~provided,
increasing
the network lifetime. 
Although works
have~proposed, for instance,  
communication protocols to ma\-ximize the network lifetime, they did not focus on 
the re\-dundancy of sensed data, transmitting 
all the collected~information.}

~\cp{Data compression techniques by local data lossy or lossless allow us to aggregate and decrease the gathered information redundancy in WBAN~\cite{giorgi2017combined}. While in lossy compression, we discard an acceptable amount of information, reducing the number of transmitted bits and increasing data compression rates, in lossless compression, the information remains unchanged~\cite{tsai2018efficient}. 
Some works have employed lossy compression by a threshold between each transmitted data~\cite{marcelloni2010enabling, azar2018performance}, while others employ codification tables, decreasing the amount of transmitted data and preserving their integrity~\cite{antonopoulos2016resource, azar2018using}.~\ca{However, as WBAN sensors collect distinct kinds of data, they require both techniques to work in a hybrid way~\cite{deepu2017hybrid,giorgi2017combined}.}
Therefore, the management of the compression service must be adaptive to the characteristics of the detected data to be energy efficient. Though many of works that take into account signal heterogeneity 
disregard the maximum latency set for WBAN to medical applications, 125ms, and to other ones, 250ms~\cite{movassaghi2014wireless}.}

~\cp{This work presents a mechanism for the management of local data compression in real-time and energy-efficient to WBAN called GROWN (\textit{Ener\textbf{G}y-Efficient Local Data Compression for T\textbf{R}ansmission \textbf{O}ver \textbf{W}BA\textbf{N}}). GROWN coordinates the compression to the type of the collected data signal, avoiding redundant data transmission and thus reducing devices’ energy consumption. GROWN applies lossy compression by a threshold to limit the acceptable data losses during transmissions, and lossless compression by encoding tables, which normalizes the data format to decrease the amount of transmitted information. Experimentation points out that GROWN increases the lifetime of sensors up to 53.73\% with a maximum 
\agnc{delay}
of 55ms between consecutive data samples, showing 
gains to WBAN working in real-time.}

The paper is organized as follows: Section~\ref{sec:rel} shows the related works. Section~\ref{sec:mec} describes the GROWN system and its operation on the WBAN.
Section~\ref{sec:eval} shows an experimental evaluation and the results obtained. Section~\ref{sec:con} presents the conclusion and future works.

\section{Related Work}
\label{sec:rel}

\cp{The literature has shown various  local and hybrid data compression techniques~\cite{giorgi2017combined},~\cite{antonopoulos2016resource,azar2018using,azar2018performance,deepu2017hybrid}.
The local data compression classes take advantages, such as reduced information redundancy and lower power consumption in wireless transmission. However, compressing data can lead to increased latency in the delivery of information, in some cases, higher energy consumption is due to the need for more computational processing for compression. Among works that have applied local data compression for WBANs, in~\cite{azar2018performance}, an amendment to the traditional lightweight temporal compression (LTC) method has been proposed by~\cite{schoellhammer2004lightweight} to minimize the error rate of information rebuilding and decrease the redundancy of the information transmitted. They also added the differential pulse coding modulation (DPCM) method into the compression phase to define these combinations according to the target application. While both methods increase the compression rates and decrease the number of transmissions, the developed method needs the storage of the data set to be transmitted over a time of 60 seconds to check the consecutive samples; this issue directly impact on information delay, usually a primary requirement for WBAN~\cite{movassaghi2014wireless}. In~\cite{antonopoulos2016resource}, a lossless method explores the correlation between consecutive samples to get an efficient table of dynamics for Huffman coding. Though, 
the method exhibited high processing rates due to the constant updates of the dynamic table, affecting the device's energy consumption. 
In~\cite{azar2018using}, the authors presented a lossless compression method based on the discrete wavelet transform (DWT) as using a lifting scheme
~\cite{sweldens1998lifting}. Thus, the transformation of redundant samples in a given period of time can be compressed with a smaller number of bits. Although the method minimized the redundancy of the information and achieved high compression rates, 
its computational cost  
requires an excessive energy consumption from the devices. The use of the local data compression classes separately offers improvements to the efficiency of the body devices. In~\cite{giorgi2017combined},
DPCM was applied for loss compression and the coding tables for Exponential of Golomb (Exp-Golomb)~\cite{teuhola1978compression} with changes for lossless compression. The method achieved high compression rates, however, the energy consumption of the devices and the compression delay is not explained. In~\cite{deepu2017hybrid},
a hybrid-search method decreases the power consumption of devices and improves the quality of the electrocardiogram (ECG) signal. The data are compressed with lossy compression, suitable for the preliminary evaluation of ECG signals; and in cases that require more detailed analysis, it uses lossless data compression. Although there is a 
considerable reduction of the power consumption of the devices, the method only classified one type of signal and disregarded information such as device latency and memory.}

\section{Real-Time Local Data Compression}
\label{sec:mec}

\cp{This section presents an overview of the collect and dissemination data environment, the GROWN components and their operation functioning. }
\agnc{GROWN decreases the amount of~transmitted information from wearable devices (e.g., sensors) to the central entity (e.g., sink), whose processing and energy resources allow operation while not affecting the latency in WBAN.}
\cp{We also assume that the sink device disseminates sensitive data collected from body devices to outside devices to enable them to follow data evolution over time.}

\begin{figure}[!b]
    \centering
    \includegraphics[width=0.9\linewidth]{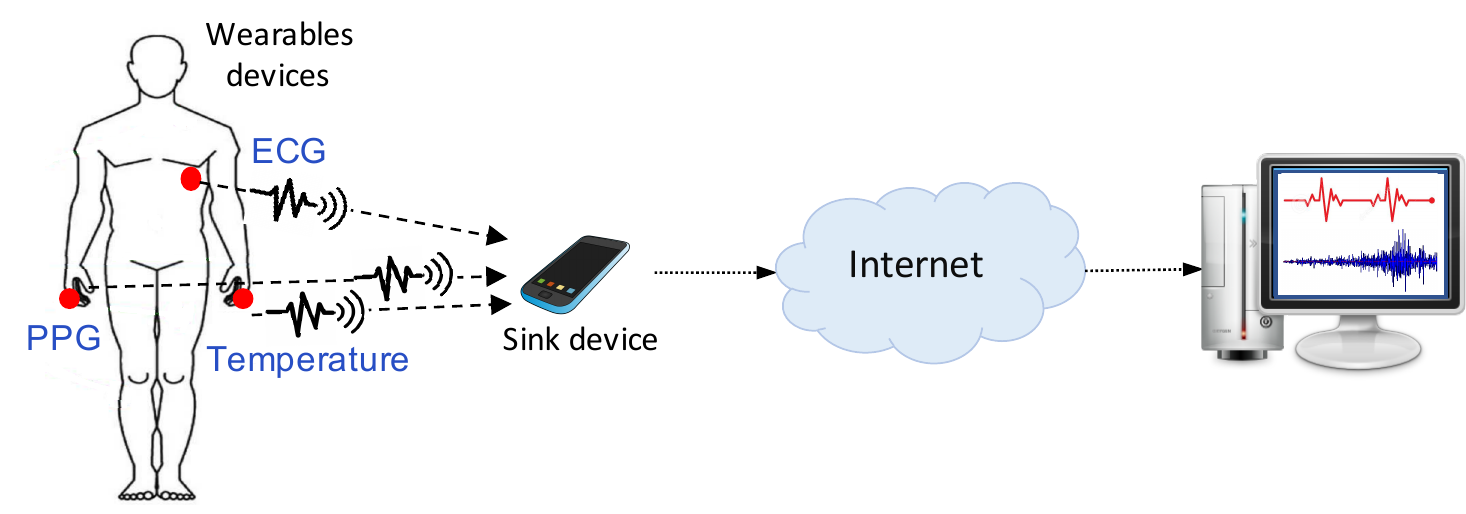}
    \vspace{-0.05cm}
    \caption{Gather and dissemination model of signals}
    \label{Fig:modeloderede}
\end{figure}

\cp{\mbox{GROWN} carries out on a set of wearable portable devices (nodes) for sensing personal physiological signals interconnected in a WBAN,
whose 
devices are denoted by \mbox{$D = \{d_1,d_2,d_3,..., d_j\}$}, where $ d_j \in D $, as shows Fig.~\ref{Fig:modeloderede}. } 
\cp{These devices possess computational resources to collect and disseminate physiological signals. All devices keep a unique identifier ($Id$)  to identify it over time and perform the tasks of signal gather and processing, as well as sending the information
to the sink device.}
\cp{By simplicity, we assume that the wireless communication technology
controls
the message losses between sensors and the sink device. Also, nodes operate statically over time in a star network topology.} 

\begin{figure}[h]
    \centering 
    \includegraphics[width=\linewidth]{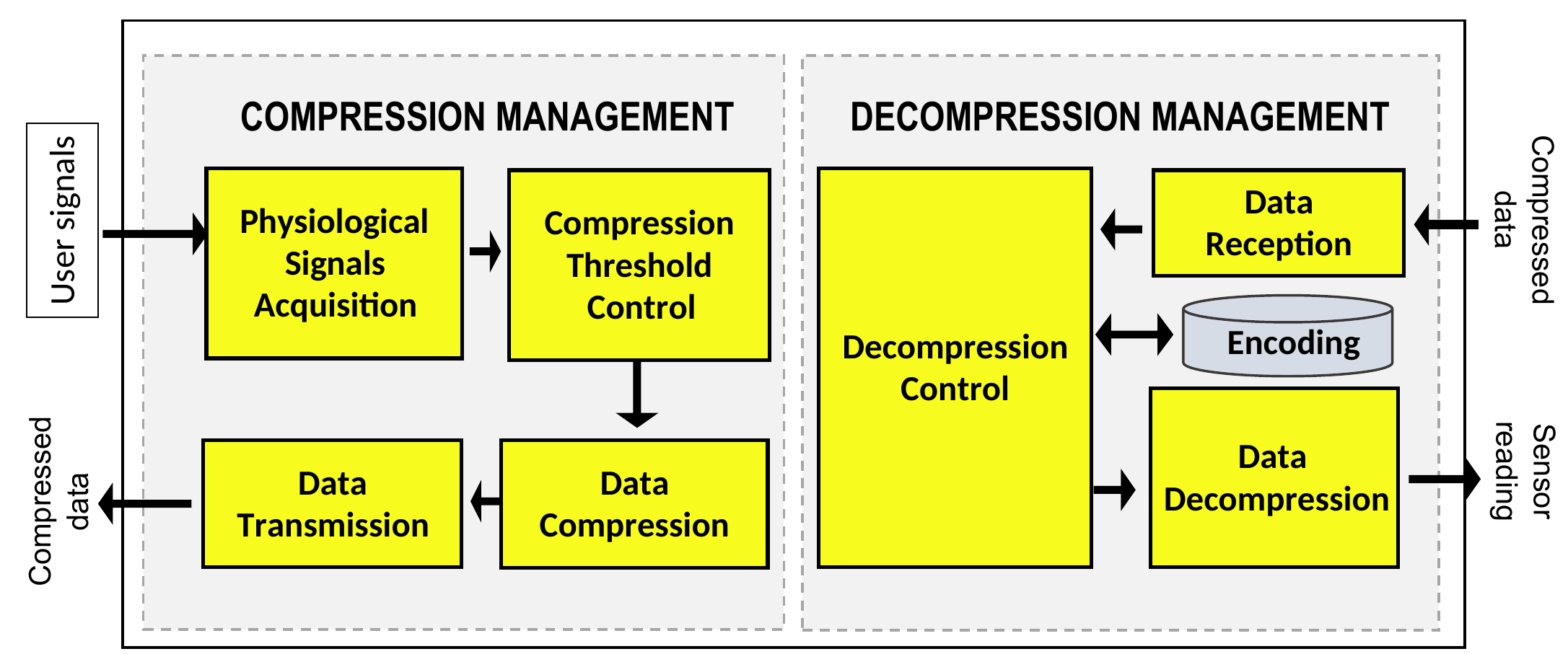}
    \caption{The GROWN architecture}
    \label{Fig:arquitetura}
\end{figure}

\vspace{0.2cm}
\cp{The GROWN architecture takes 
the {\bf Compression Management (CM)} and {\bf Decompression Management (DM)} modules, as shown in Fig.~\ref{Fig:arquitetura}. The former 
 makes the compression of 
the physiological signals received from wearable devices and the latter decompresses and recovers the original information collected from sensors. We detail each one as follows.}

\vspace{-0.2cm}
\noindent

\subsection{Data Compression Module}
\cp{The CM module consists of components that convert the physiological signal from analog to digital, analyze the digital data and define the form of compression, perform data compression and transmit that data to the sink device through a wireless connection for further decompression. Wearable devices collect physiological signals from the user through the
\textit{Physiological Signals Acquisition} ($PSA$) component. This component converts the analog physiological signal to digital and sends it to the \textit{Compression Threshold Control} ($CTC$) component. As described in Algorithm~\ref{alg:compression}, this component receives the current reading from the sensor ($l$.1). The first reading of the sensor ($l$.3) is then stored for comparison with the next ones ($l$.4), and also sent for compression ($l$.6.14). In the next readings, if the variation between the current and the previous reading exceeds the previously established threshold ($l$.8-9), the difference is forwarded to compression ($l$.11)
\agnc{to decrease the amount of transmitted information,}
and the last reading stored for the next comparison ($l$.12). }

\begin{algorithm}[!htb]
{
\setstretch{0.85}
{\fontsize{8}{10}
\textbf{procedure} \selectfont{\textsc{CompressionControl}}$ \, (actual\_reading)$\\
\sep $compression\_value \leftarrow 0$\\
\sep \textbf{if} $(first\_reading == True)$\\
\sep \sep $last\_reading \leftarrow actual\_reading$\\
\sep \sep $first\_reading \leftarrow False$\\
\sep \sep $compression\_value \leftarrow actual\_reading$\\
\sep \textbf{else}\\
\sep \sep $aux\_reading \leftarrow |last\_reading - actual\_reading|$\\
\sep \sep \textbf{if} $((t_{threshold} > 0$ AND $ aux\_reading > t_{threshold})$ OR\\
\sep \sep \sep \sep $(t_{threshold} == 0))$\\ 
\sep \sep \sep \sep $compression\_value \leftarrow (actual\_reading - last\_reading)$\\
\sep \sep \sep \sep $last\_reading \leftarrow actual\_reading$\\
\sep \sep \textbf{endif}\\
\sep \textbf{endif}\\
\sep \textbf{return} $(compression\_value)$\\
\textbf{end procedure}\\
\caption{Compression control}
\label{alg:compression}
}}
\end{algorithm}

\cp{
GROWN compresses the information ($e_{ts}$) received by $CTC$
using a modified version of the Exponential-Golomb code (Exp-Golomb) of order 3~\cite{teuhola1978compression}. Thus, each received sample ($e_{ts} \neq 0$) is represented as a bit sequence ($bs_i$) composed of two parts  $s_i | a_i$. The first part ($s_i$) identifies the group to which $e_{ts}$ belongs and illustrates the value of $n_i$, which is equivalent to the number of bits needed to represent $e_{ts}$.
Thus, we represent the groups $s_i$ related to the first $2^k - 1$ values of $n_i$
by $\left \lfloor \log_2(|e_{ts}|) \right \rfloor +1$, where~$s_i$ is represented by $k$ bits. 
When $n_i > 6$, the value of the $s_i$ group is determined by $\left \lfloor \log_2(|e_{ts}|) - 1 \right \rfloor$, where the first $s_i - 1$ values of $s_i$ are represented by $1$ followed by $0$~(zero). The $a_i$ part corresponds to the binary representation of $e_{ts}$, 
whose variable-length code is defined according to~\cite{marcelloni2008simple}, ensuring different values for the entries of each group. Table~\ref{tab:regras} shows the rules for $e_{ts}$ values. }

\begin{table}[h]
\centering
\caption{Data compression rules}
\label{tab:regras}
{ \small
\begin{tabular}{m{1.2cm}||lm{2.3cm}}
\hlineB{2}
\textbf{Condition} & \textbf{Rule} \\ \hlineB{2}
$e_{ts} < 0$ & Make a 2 complement of $e_{ts}$, subtract 1 and use\\
& the least significant $n_i$ bits \\
$e_{ts} = 0$ & Encode $s_i$ as 000 and not represent $a_i$\\
$e_{ts} > 0$ & $a_i$ corresponds to the less significant $n_i$ bits of\\
& the $e_{ts}$ 2 complement \\
\hlineB{2}
\end{tabular}}
\end{table}

\cp{Table~\ref{tab:tabproposta} shows the number of $n_i$ bits
for representing
~$e_{ts}$ and the encoding values for each group $s_i$.
Thereby, enconding
depends on differences between the values of $e_{ts}$, thus more frequent differences imply shorter code. After $e_{ts}$ compression, the new information ($Inf_{ts}$) is transmitted to the sink device.} 

\begin{table}[h] 
\centering
\caption{Encoding values}
    \label{tab:tabproposta}
\small
\begin{tabular}{cllccrrlcc}
\hlineB{2}
\multicolumn{1}{c||}{\textbf{$n_i$}} & 
\multicolumn{1}{c}{\textbf{$s_i$}} & 
\multicolumn{1}{c}{\textbf{$e_{ts}$}} & 
\multicolumn{1}{c}{\textbf{bit}} & 
\multicolumn{1}{l}{\textbf{byte}} \\ \hlineB{2}

\multicolumn{1}{l||}{0}  & 
\multicolumn{1}{r}{000} & 
\multicolumn{1}{r}{0}  & 
\multicolumn{1}{c}{3}   & 
\multicolumn{1}{c}{1}  & \\

\multicolumn{1}{l||}{1}           & 
\multicolumn{1}{r}{001}         & 
\multicolumn{1}{r}{-1,+1}      & 
\multicolumn{1}{c}{4}            & 
\multicolumn{1}{c}{1}   \\ 

\multicolumn{1}{l||}{2}           & 
\multicolumn{1}{r}{010}         & 
\multicolumn{1}{r}{-3,-2,+2,+3} & 
\multicolumn{1}{c}{5}            & 
\multicolumn{1}{c}{1}            \\

\multicolumn{1}{l||}{3}           & 
\multicolumn{1}{r}{011}         & 
\multicolumn{1}{r}{-7,\dots,-4,+4,\dots,+7}         & 
\multicolumn{1}{c}{6}            & 
\multicolumn{1}{c}{1}            \\

\multicolumn{1}{l||}{4}           & 
\multicolumn{1}{r}{100}         & 
\multicolumn{1}{r}{-15,\dots,-8,+8,\dots,+15}       & 
\multicolumn{1}{c}{7}            & 
\multicolumn{1}{c}{1}            \\ 

\multicolumn{1}{l||}{5}           & 
\multicolumn{1}{r}{101}         & 
\multicolumn{1}{r}{-31,\dots,-16,+16,\dots,+31}     &
\multicolumn{1}{c}{8}            & 
\multicolumn{1}{c}{1}             \\

\multicolumn{1}{l||}{6}           &
\multicolumn{1}{r}{110}         & 
\multicolumn{1}{r}{-63,\dots,-32,+32,\dots,+63}     &
\multicolumn{1}{c}{9}            &
\multicolumn{1}{c}{2}             \\

\multicolumn{1}{l||}{7}           & 
\multicolumn{1}{r}{11110}        & 
\multicolumn{1}{r}{-127,\dots,-64,+64,\dots,+127}   &
\multicolumn{1}{c}{12}           & 
\multicolumn{1}{c}{2}             \\ 

\multicolumn{1}{l||}{8}           &
\multicolumn{1}{r}{111110}      & 
\multicolumn{1}{r}{-255,\dots,-128,+128,\dots,+255} &
\multicolumn{1}{c}{14}           & 
\multicolumn{1}{c}{2}             \\

\multicolumn{1}{l||}{9}           &
\multicolumn{1}{r}{1111110}      & 
\multicolumn{1}{r}{-511,\dots,-256,+256,\dots,+511} &
\multicolumn{1}{c}{16}           & 
\multicolumn{1}{c}{2}             \\ \hlineB{2}
\end{tabular}%
\end{table}

\vspace{-0.25cm}
\subsection{\agnb{Decompression management module}}
\label{subsec:dcmod}

\agn{This module plays in the sink device, which has enough resources of 
power, storage, and processing to carry out the decompression of data sent by wearable devices
in 
the WBAN. 
Thus, \agn{as depicted in Fig.~\ref{Fig:arquitetura},} the recovery of each physiological signal starts in the \textit{Data Reception} component. This component receives the message with the compressed data ($bs_i$) and the device identification ($D_{id}$), and forwards it to the \textit{Decompression Control} (DC) component, according to Algorithm~\ref{alg:decompression}. 
DC
\agnb{holds}
a reference list (RL) to store the identification of all wearable devices in 
the WBAN, besides the values of their last reading ($\hat{x}_{ts}$) \agn{($l$.2)}.
Next, CD obtains the prediction error ($e_{ts}$) through the prefix $s_i$ ($l$.3) and the suffix $a_i$ ($l$.4) from $bs_i$. Then, it queries the \textit{Encoding} database ($l$.5), which stores \agnb{the values of} indexes ($n_i$), groups ($s_i$) and
\agnb{the binary representation of $e_{ts}$}
($a_i$).
As $a_i$ is unique in $s_i$, the recovery of the received $bs_i$ occurs with few instructions, adapting to the real-time operation. Thus, DC forwards $\hat{x}_{ts}$ and $e_{ts}$ values to the
\textit{Data Decompression} (DD) component and updates the RL. Lastly,}
\agnb{DD decompress the data and makes it available to others applications.}

\begin{algorithm}[!t]
{
\setstretch{0.85}
{\fontsize{8}{10}
\textbf{procedure} \selectfont{\textsc{DecompressionControl}}$ \, (compressed\_data,id)$\\
\sep $last\_reading \leftarrow getLastReading(id)$\\
\sep $prefix \leftarrow getPrefix(compressed\_data)$\\
\sep $suffix \leftarrow getSuffix(compressed\_data)$\\
\sep $data\_value \leftarrow bdcoding(prefix,suffix)$\\
\sep \textbf{return} $(data\_value,id)$\\
\textbf{end procedure}\\
\caption{Decompression control}
\label{alg:decompression}
}}
\end{algorithm}

\subsection{Operation}

\begin{figure*}[!htb]
    \centering
    \includegraphics[width=0.95\linewidth]{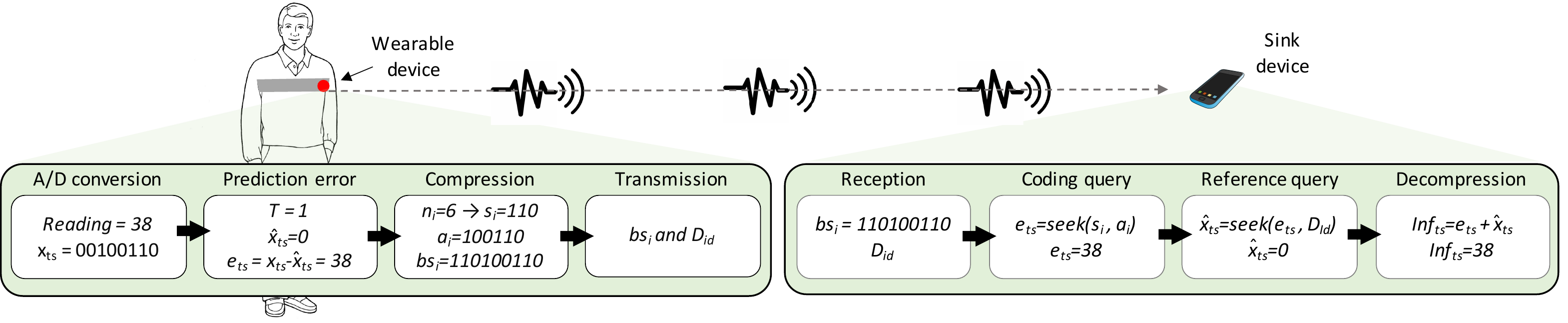}
    \caption{The step by step of GROWN operation}
    \label{Fig:exemplofunc}
\end{figure*}

\agn{Fig.~\ref{Fig:exemplofunc} depicts a wearable device in a WBAN continuously sensing a person's temperature, meanwhile it interacts with the sink device
(e.g., smartphone). The sensor starts collecting the temperature signal and converts it from analog to digital in a binary representation ($x_{ts}$) with $R$ bits. Suppose its reading is $38$, so $x_{ts}$ value in $8$ bits is $001001100$. As the decision-making about the data compression relies on a pre-defined prediction filter $T$, assume $T = 1$. Hence, the filter checks the $x_{ts}$ value and as this is the first reading, the last reading ($\hat{x}_{ts}$) gets the value $0$. Thus, the prediction error ($e_{ts}$), which means the difference between consecutive readings, is equal to $e_{ts}=x_{ts} - \hat{x}_{ts} = 38$.}

\agn{Next, the $e_{ts}$ value is encoded to a bit sequence ($bs_i$) in two parts, $s_i|a_i$. $s_i$ represents the group ($n_i$) which $e_{ts}$ belongs, so $n_i = \left \lfloor \log_2(|38|) \right \rfloor + 1 = 6$. By the encoding values in Table~\ref{tab:tabproposta}, we verify that the $s_i$ value belongs to the group $110$. The $a_i$ value comes from the conversion of the $e_{ts} = 38$ value to a binary form, $a_i = 100110$, so $bs_i = 110100110$. After that, the wearable device sends a message to the sink device carrying $bs_i$ and $D_{Id}$ values. When the sink receives this message, it recovers the $s_i$ and $a_i$ values from $bs_i$ and gets the prediction error $e_{ts}$ by querying the encoding table using $bs_i$ and $D_{Id}$ values. Through RL, it obtains $\hat{x}_{ts}$ from $e_{ts}$ and $D_{Id}$. As it is the first reading, $\hat{x}_{ts} = 0$. Finally, it retrieves the information equivalent to the reading of the wearable device, $Inf_{ts} = e_{ts} + \hat{x}_{ts} =~38$.}

\section{Evaluation}
\label{sec:eval}

This section describes the GROWN implementation and analyzes how the management of lossless (LL) and lossy (LS) compression provides energy efficiency to WBAN.
GROWN combines distinct platforms to meet the needs of wearable and sink devices to mimics a real environment in experimentation. We have compiled the compression module
in C++ language in Arduino boards, whose
integrated microcontroller operates with a 16MHz clock and a 10 bits resolution in each logical port for physiological sensors readings.
\agnb{We have developed
the decompression module
in Android Studio, version 3.6.1, that was installed in the sink device, a Motorola smartphone, model G$^4$ Plus, with an Android version~7.0.} 

We built a testbed composed by three wearable devices to act as physiological sensors and one smartphone to operate as the WBAN's sink device, as depicted in Fig.~\ref{Fig:modeloderede}. 
We set up \agnb{the wearables devices} on protoboards model Mb102 to sense temperature, ECG and photoplethysmogram (PPG) signals, as
described in Table~\ref{tab:tabdispositivos}. These protoboards provide energy to sensors and Arduino boards supplied by 5V regulated power sources~supported by 9V/400mAh batteries. Each wearable device has a sensor to collect a given  physiological signal and all the sensors codes were compiled with official libraries.

\begin{table}[!b]
\centering
        \caption{Wearable devices settings}
        \label{tab:tabdispositivos}
        \relsize{-1.7}
        \begin{threeparttable}
        \setlength{\tabcolsep}{2pt}
        \renewcommand*{\arraystretch}{1.5}
        \setlength{\extrarowheight}{1.0pt}
        \begin{tabular}{l||ccc}
        \hlineB{2}
        \multirow{2}{*}{\textbf{Feature}}&\multicolumn{3}{c}{\textbf{Device}} \\ \cline{2-4}
        & \textbf{Temperature}& \textbf{PPG}& \textbf{ECG / ECGDB}\\ \hlineB{2}
        \textbf{Board} & Arduino UNO &Arduino UNO&Arduino Mega\\ 
        \textbf{Controller}&ATmega328P&ATmega328P&ATmega2560\\ 
        \textbf{Body sensor} & MLX90614&PulseSensor& AD8232\\ \hlineB{2}
        \end{tabular}
        \end{threeparttable}
\end{table}

\begin{figure}[h] 
    \centering
    \includegraphics[width=0.8\linewidth]{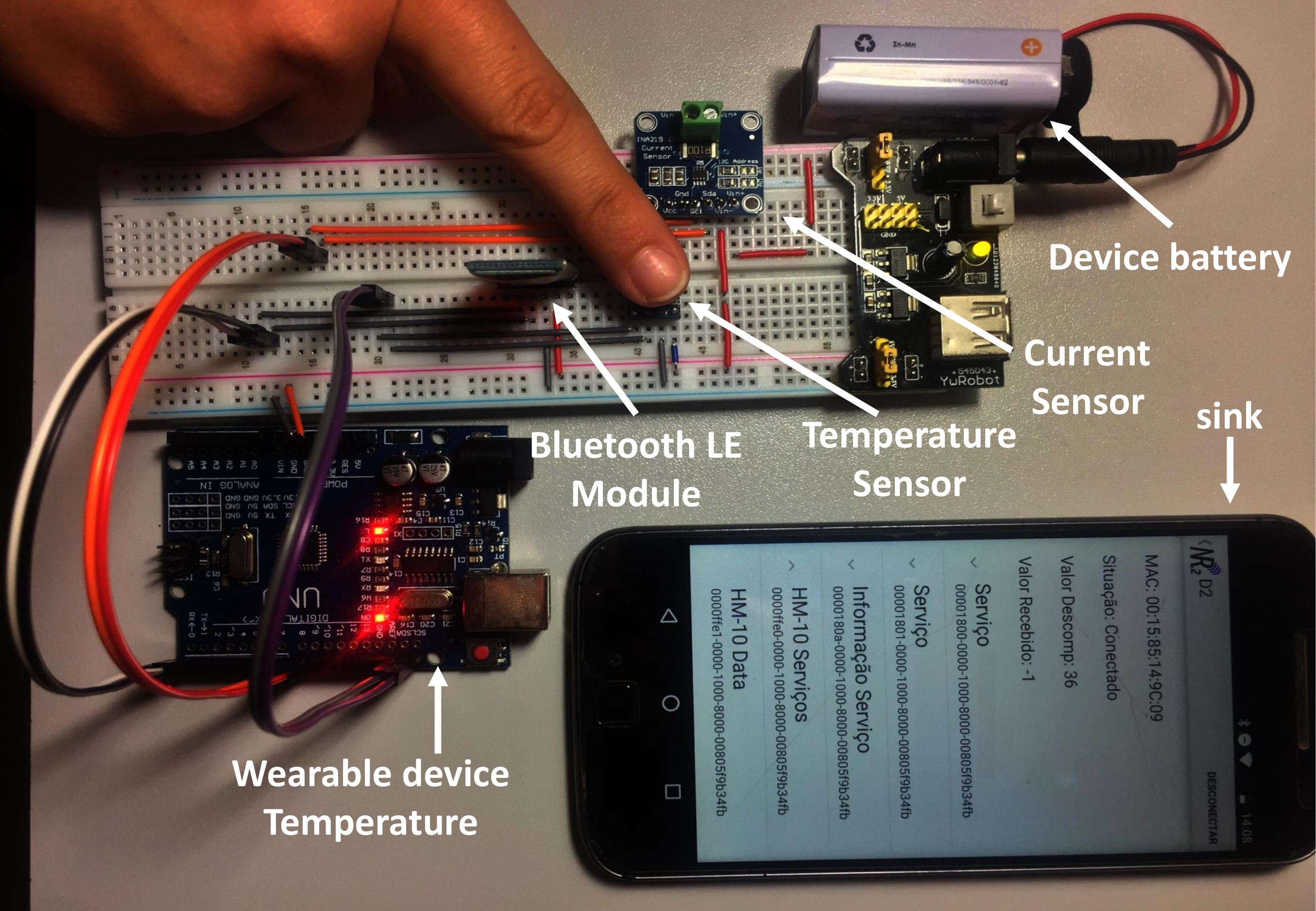}
    \captionof{figure}{\agnb{Wearable temperature device testbed}}
    \label{Fig:dispositivoTemperatura}
\end{figure}

\agn{To
\agnb{enhance}
the GROWN analysis in the ECG signals compression, we have
\agnb{added}
a 4th device (ECGDB) to our testbed that acts as an ECG sensor and employs the MIT-BIH arrythmia database~\cite{mitdb2005} as its data source. The wearables and sink devices communicate by Bluetooth Low Energy (BLE), \agnb{by IEEE 802.15.1 standard},
\agnb{through a}
HM-10 BLE V4
module. 
\agnb{They keep a connection}
during 600s, a sufficient period to get 600 instantaneous current reading samples from each device. We set up the smartphone to operate in flight mode during the experiments to avoid any interruptions from other applications, while keeping in operation only the Bluetooth communication. As shown in Fig.~\ref{Fig:dispositivoTemperatura}, the environment for sensing a person's body temperature operates by infrared signals to avoid the physical contact of the person's body with the device. After the reading, the 
device compresses such data and forwards to the Bluetooth module that transmits it to the sink device, which then decompresses the data. We have set up the prediction filter to 1 (T = 1) for lossy compression to improve the identification of the correlation between consecutive samples.}

\agn{The analysis of GROWN 
takes into account 
the metrics presented in Table~\ref{tab:metricas}.
We have measured the~\textit{Compression Delay (CD)} and~\textit{Decompression Delay~(DD)} by~\cite{patel2013simulation}, and the~\textit{Average Delay (AD)} by~\cite{zordan2014performance}.
We have also employed a current and voltage sensor from Texas Instruments, model INA219~\cite{marcelloni2010enabling} to measure the~\textit{Device Energy Consumption (DEC)}. This sensor monitors the voltage drop over a~\SI{1}{m\ohm} shunt resistor, which is proportional to the current flowing through it~\cite{parks2007ohms}.
\agnd{We obtained the~\textit{Compression Ratio (PCR)} from the amount of compressed packets ($comp\_pkt$) and the amount of these packets without any compression ($orig\_pkt$).}
Results exhibited correspond to an average of ten repetitions.}

\begin{table}[h] 
\renewcommand*{\arraystretch}{1.5}
\centering
\caption{Evaluation Metrics}
\label{tab:metricas}
{ 
\footnotesize
\begin{tabular}{l||l}
\hlineB{2}
\textbf{Metric} & \textbf{Equation} \\ \hlineB{2}

\textbf{\textbf{Compression Delay (CD)}} 
& $\sum\limits_{i=1}^{n}\frac{T_{fcom}i - T_{icom}i} {Inf_{ts}i}$ \\ \hline

\textbf{\textbf{Decompression Delay} (DD)}& $\sum\limits_{i=1}^{n}\frac{T_{fdes}i - T_{icdes}i} {Inf_{ts}i}$\\ \hline

\textbf{Average Delay (AD)} & $\sum\limits_{x\;=\;1}^{N_e} \sum\limits_{y\;=\;1}^{T_{tr}}  \frac{CD_{xy}+DD_{xy}+DTR_{xy}}{T_{tr} \; \times \; N_e}$ \\ \hline
\textbf{Energy consumption (DEC)}& $I \times t$ \\ \hline

\textbf{\textbf{Compression Ratio} (PCR)} & $\left (1- \frac{comp\_pkt}{orig\_pkt} \right ) \times 100$ \\ 
\hlineB{2}
\end{tabular}
}
\end{table}

\begin{figure*}[!t]
    \centering   
    \includegraphics[width=0.24\linewidth]{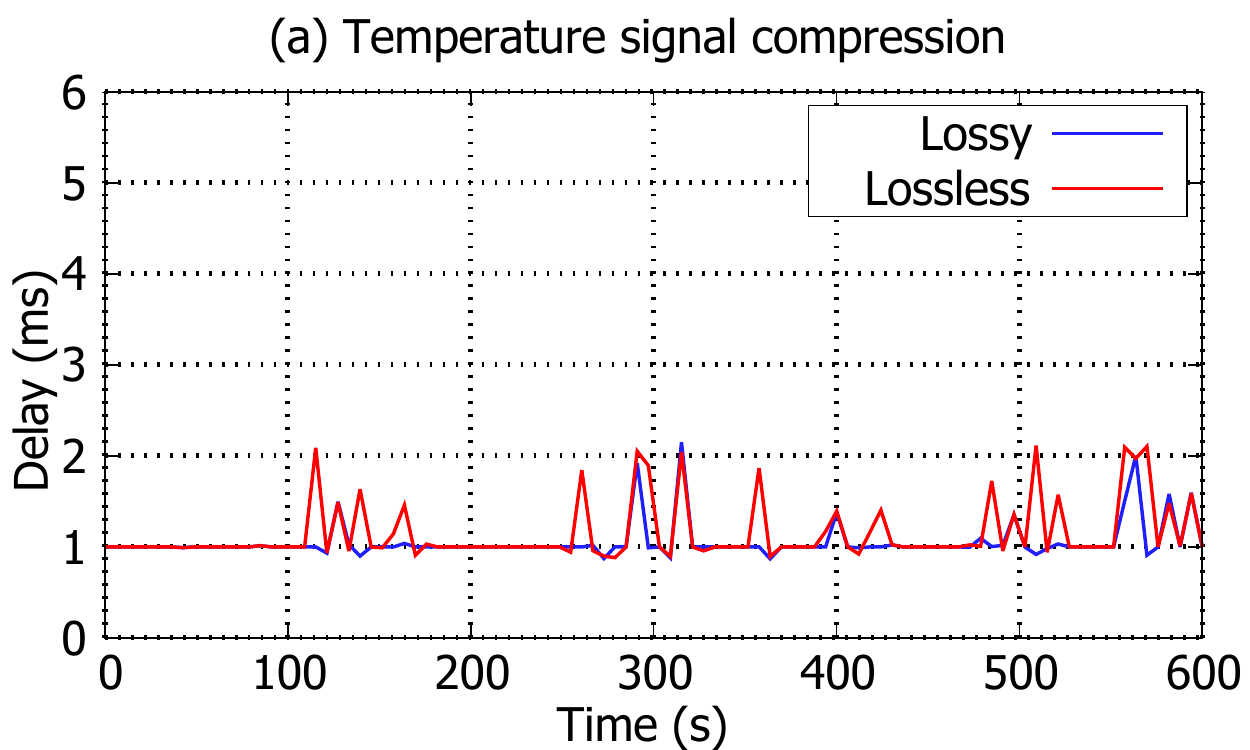}
    \includegraphics[width=0.24\linewidth]{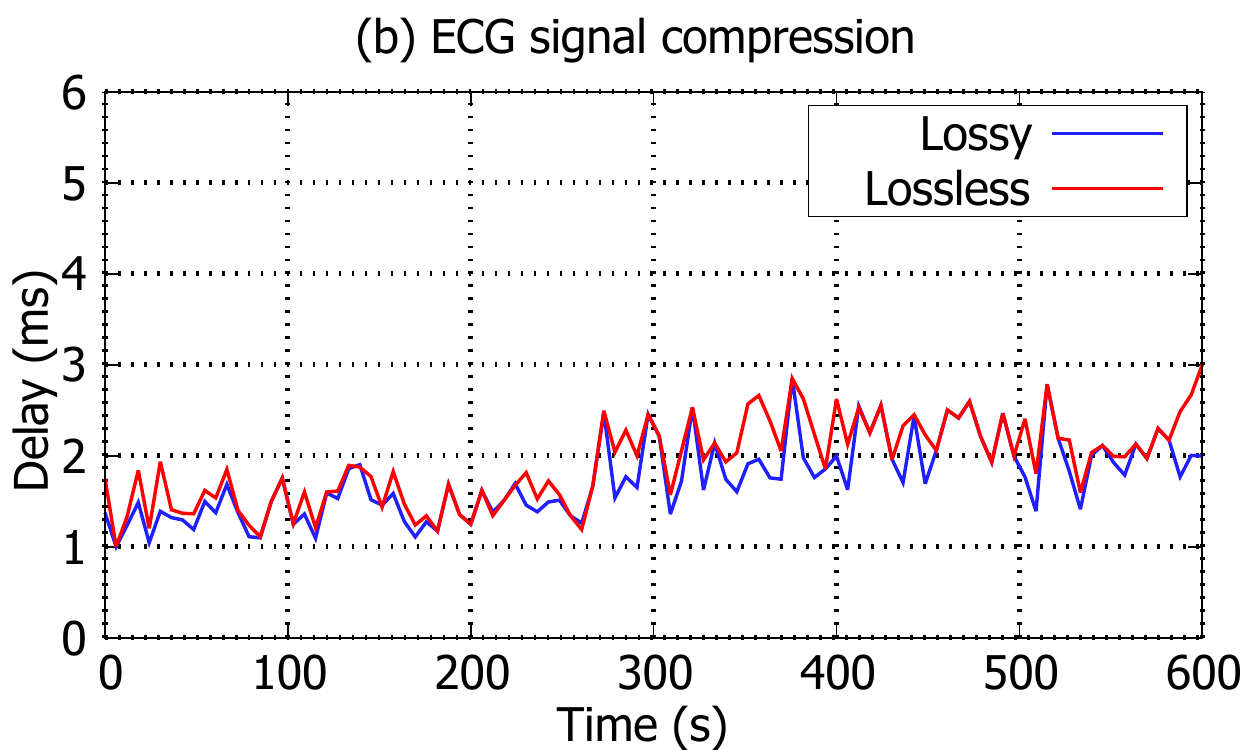}
    \includegraphics[width=0.24\linewidth]{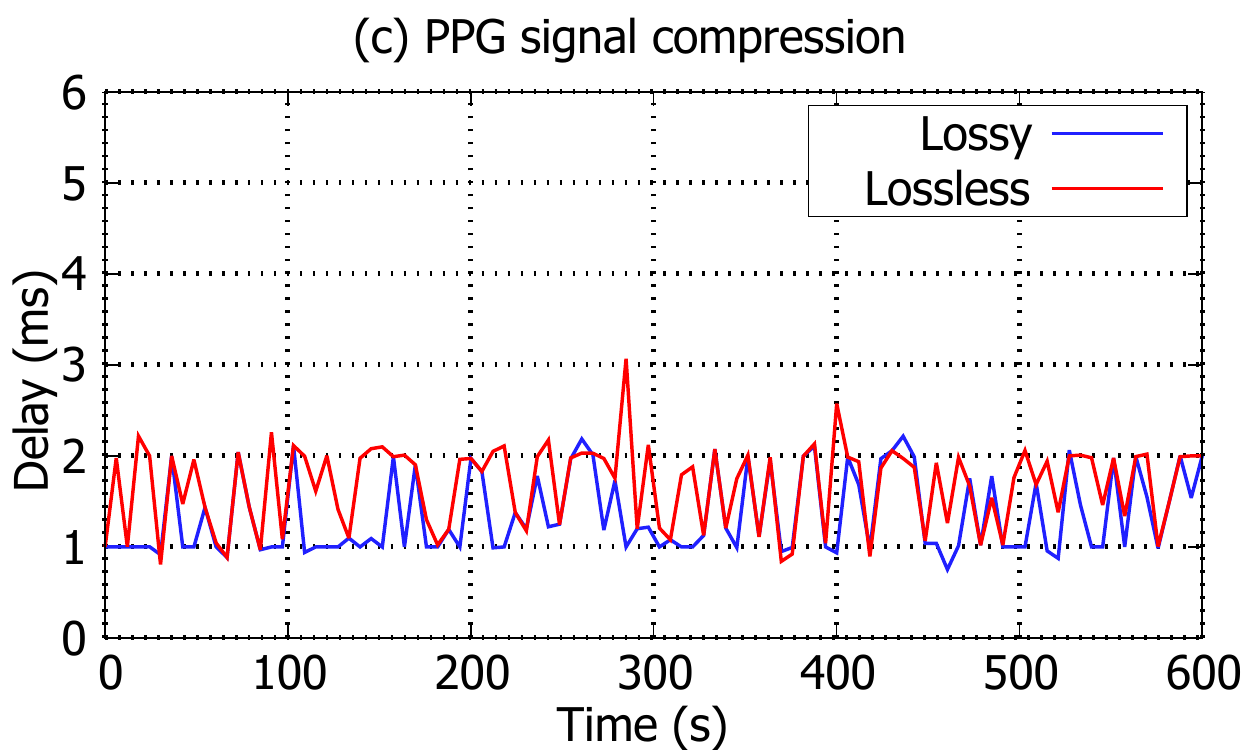}
    \includegraphics[width=0.24\linewidth]{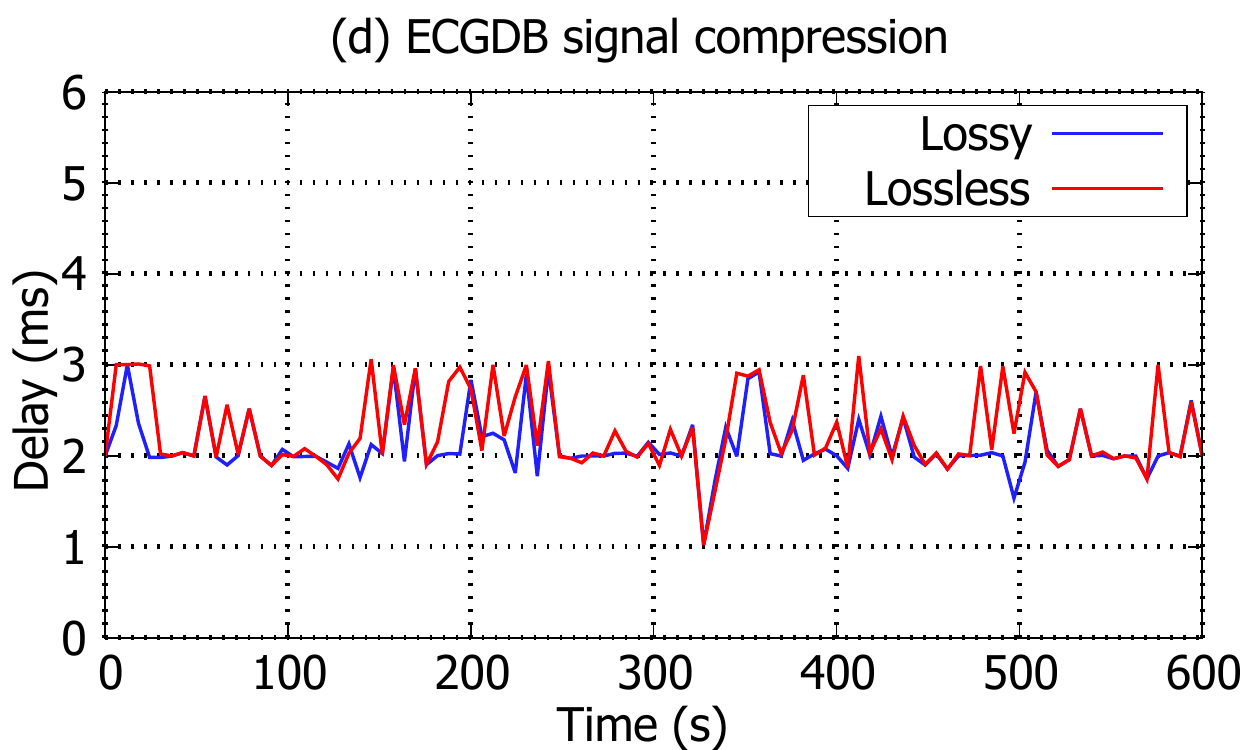}
    
    \vspace{0.2cm}
    
    \includegraphics[width=0.24\linewidth]{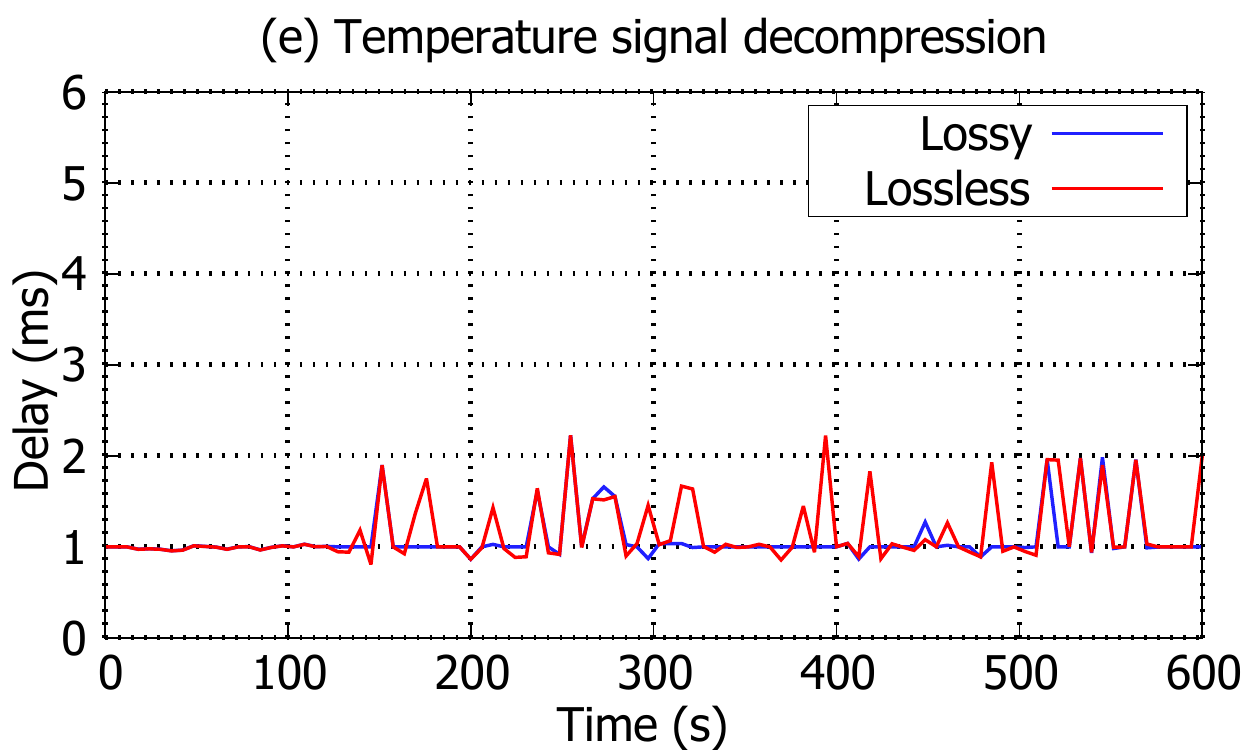}
    \includegraphics[width=0.24\linewidth]{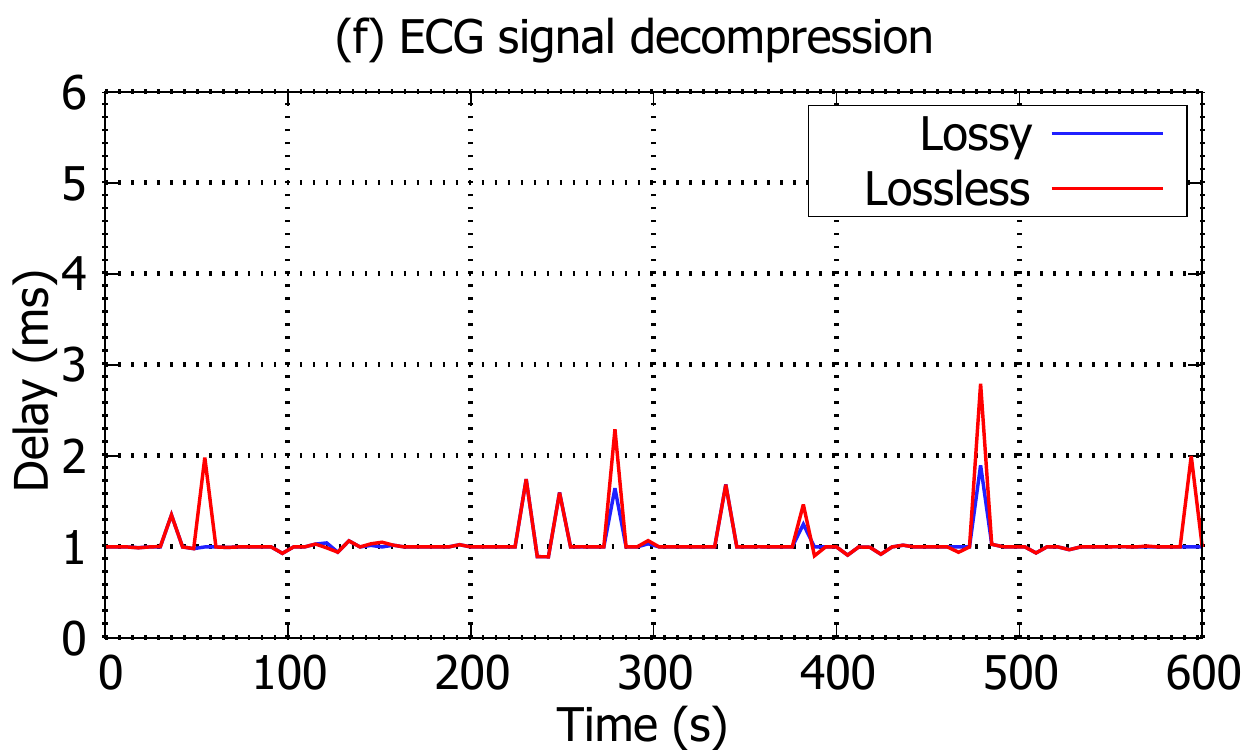}
    \includegraphics[width=0.24\linewidth]{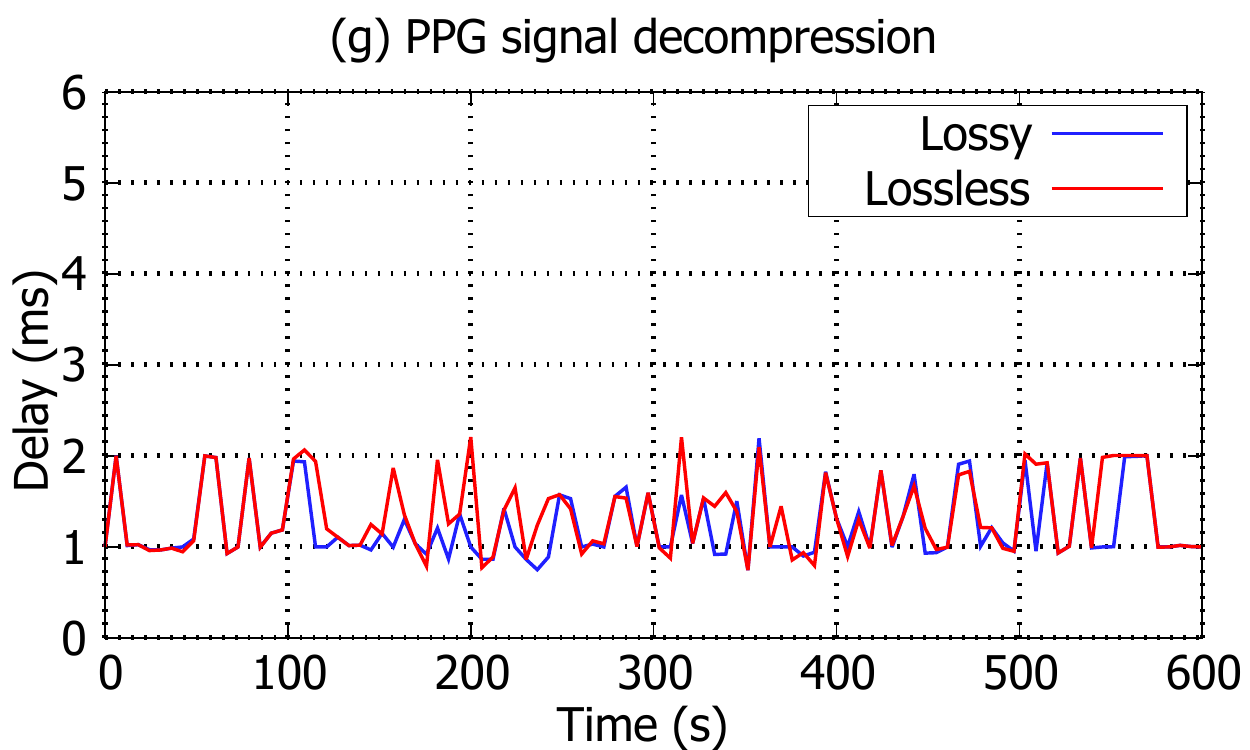}
    \includegraphics[width=0.24\linewidth]{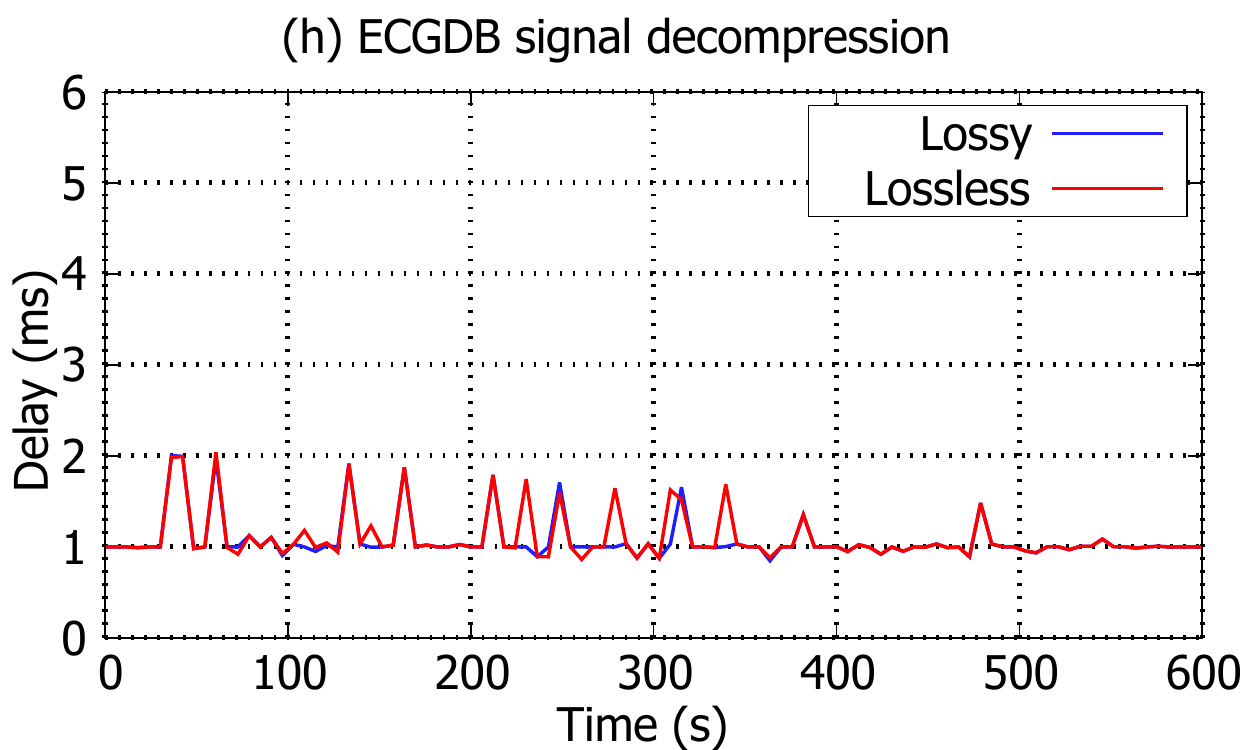}
    \caption{Delays in compression and decompression of physiological signals}
    \label{Fig:TempoComDes}
\end{figure*}

\vspace{0.1cm}
\subsection{Results} 
\label{sec:analysis}

\agn{The compression and decompression processes
\agnb{add a delay to}
data dissemination, which varies according to the correlation between consecutive samples.
%
\agnd{The}
temperature signals exhibit high correlated samples,
\agnd{whose consecutive values present slight variations. Hence,}
GROWN measured a value of $\approx1$ms to CD, the lowest among four types of signals~(Fig.~\ref{Fig:TempoComDes}(a)). 
We noted that the samples correlation of ECG (Fig.~\ref{Fig:TempoComDes}(b)) and PPG (Fig.~\ref{Fig:TempoComDes}(c)) \agnb{signals} frequently varies over time, so
\agnb{both suffered}
higher CD, $\approx 3$ms and $\approx 2$ms, respectively. 
Finally, as shows Fig.~\ref{Fig:TempoComDes}(d),
ECGDB
\agnd{signals behaved in similar way}
and
\agnb{cost}
$\approx 3$ms for CD. On the other hand, \agnc{as depict Fig.~\ref{Fig:TempoComDes}(e), \ref{Fig:TempoComDes}(f), \ref{Fig:TempoComDes}(g), and \ref{Fig:TempoComDes}(h)},
the GROWN cost was $\approx 1$ms for DD, as
\agnb{the smartphone that plays as sink device}
has greater computational resources than wearables devices.}

\begin{figure}[h] 
    \centering 
    \includegraphics[width=0.8\linewidth]{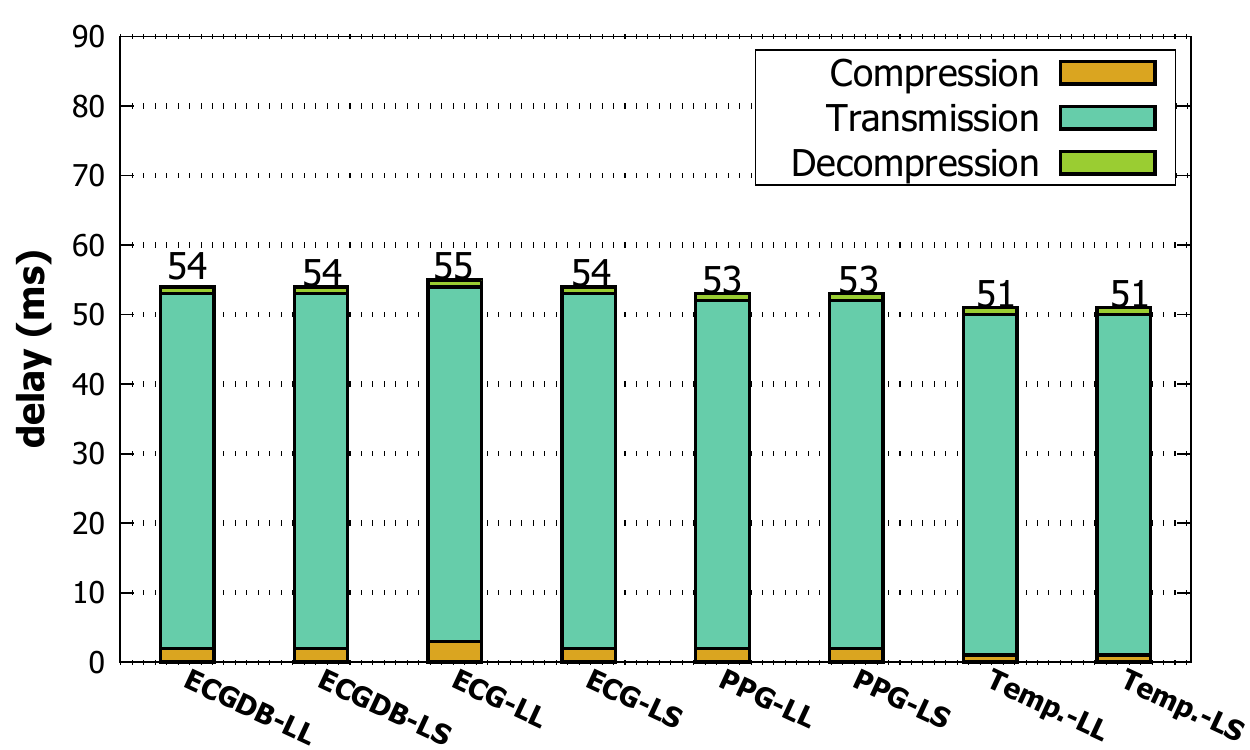}
    \caption{\agnb{Delay in processing physiological signals}}
    \label{fig:am}
\end{figure}

We have identified that \agnb{signals with} low correlated consecutive samples \agnd{(i.e., consecutive values exhibit wide variations)} increase the compression time in devices with constrained resources.
Fig.~\ref{fig:am} shows the Average Delay (AD)
\agnb{measured}
on the
\agnb{evaluated physiological signals}
with lossless (LL) and lossy (LS) compression. We noted  that ECG, ECGDB and PPG
\agnb{signals}
show $\approx 3$ms of AD, while \agnb{the} temperature signal shows $\approx 1$ms. However, the delay~of $\approx 49$ms on all wearable devices is due to the manner Bluetooth BLE connections happen, endorsing  what Gatouillat et al.~in \cite{gatouillat2018building}  have already identified on Bluetooth connections on Android 7.0. In a general manner GROWN raised 7.84\% on the total transmission time for ECG and ECGDB signals, 6\% for PPG, and 2.04\% for temperature signal. 
Those values show GROWN's ability in
saving energy on data dissemination  in 
WBANs, which demand a maximum latency of 125ms for medical applications and 250ms for non-medical
ones~\cite{movassaghi2014wireless}.

\begin{figure}[!h]
    \centering 
    \includegraphics[width=0.75\linewidth]{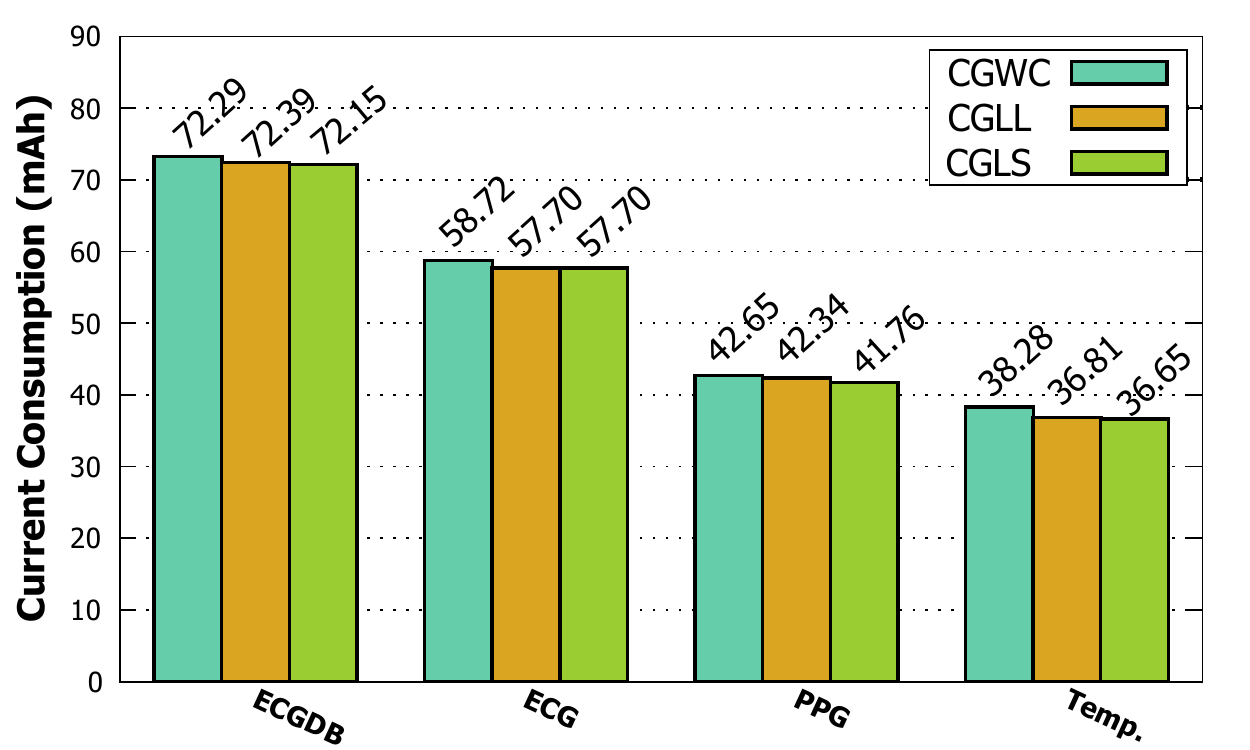}
    \caption{Energy consumption of wearables devices}
    \label{fig:ced}
\end{figure}

\agnb{We also noted that consecutive highly correlated samples enable devices to reduce energy consumption~(DEC),
as shows Fig.~\ref{fig:ced}, because GROWN assigns shorter codes for the small differences to be transmitted.}
\agnb{To verify
\agnb{its}
energy efficiency,}
we have conducted experiments
with lossless (CGLL) and lossy (CGLS) compression, and also
without compression (CGWC).
\agnd{We verified that the}
temperature device consumes on average 36.81mAh for CGLL and 36.65mAh for CGLS, while in CGWC it consumes 38.28mAh. Hence, GROWN achieves an energy efficiency of 1.47mAh for CGLL and 1.63mAh for CGLS. GROWN reached an energy efficiency of 1.02mAh on the ECG device for both methods. For the PPG device, we observe an energy efficiency of 0.31mAh for CGLL and 0.89mAh for CGLS, as this kind of signal presents low correlated consecutive samples. As the ECGDB device has performed
\agnb{frequent}
readings of ECG samples from the MIT-BIH database in the SD module, this approach has increased its energy consumption to 73.29mAh in CGWC, and produces an energy efficiency of 0.90mAh for CGLL and 1.14mAh for CGLS. 
\agnd{Although data compression also consumes energy, 
even so there is an increase of
the lifetime of each wearable device with
the use of GROWN,
as we can see in
Table~\ref{tab:tablifetime}.}

GROWN
enables an
energy efficiency to the devices even \agnb{with their}
Bluetooth module active and paired with the
\agnb{smartphone.}
But, only this condition already increases the energy consumption, as this module carries a CC2540 BLE chip, which consumes 24mA to data transmission and reception~\cite{CC2540TI2020}. Hence, to reduce energy consumption, we have set up the Bluetooth module in sleep mode. So, it falls asleep after a certain period and wakes up with a proper command. We have set up the temperature device to this configuration, as it has the highest correlated consecutive samples and it is active whenever the device does not transmit two consecutive samples. 
This strategy has reduced DEC for the temperature device to 25.81mAh for CGLL and to 24.92mAh for CGLS, both from CGWC.
Hence, as shows Table~\ref{tab:tablifetime}, it 
has increased temperature device lifetime by 53.73\% for CGLL, from 10.45h in normal mode to 15.50h in sleep mode, 
and by 48.37\% for CGLS,
from 10.45h in normal mode to 16.05h in sleep mode.

\begin{table}[!t]
        \centering
        \caption{Lifetime of wearable devices}
        \label{tab:tablifetime}
        \relsize{-1.7}
        \begin{threeparttable}
        \setlength{\tabcolsep}{2pt}
        \renewcommand*{\arraystretch}{1.0}
        \setlength{\extrarowheight}{1.0pt}
        \begin{tabular}{ll||cccc|c}
        \hlineB{2}
        \multicolumn{2}{l||}{\textbf{Mode}} & \multicolumn{4}{c|}{\textbf{Normal}} & \textbf{Sleep}    \\ \hline
        \multicolumn{2}{l||}{\textbf{Device}} & ECGBD & ECG & PPG & Temperature & Temperature\\ \hlineB{2}
        \multicolumn{1}{c|}{\multirow{3}{*}{\textbf{Battery life} ($h$)}} & CGWC  &  5.46 & 6.81 & 8.38 & \textcolor{blue}{10.45} & -  \\
        \multicolumn{1}{c|}{}                     & CGLL & 5.53 & 6.93 & 9.45 & 10.87 & \textcolor{blue}{15.50} \\
        \multicolumn{1}{c|}{}                     & CGLS & 5.54 & 6.93 & 9.58 & 10.91 & \textcolor{blue}{16.05} \\ \hlineB{2}           
        \end{tabular}
        \end{threeparttable}
\end{table}

\begin{table}[h] 
\centering
\renewcommand*{\arraystretch}{1.0}
\relsize{-1}
\begin{threeparttable}
\setlength{\tabcolsep}{3pt}
\caption{Compression ratio of physiological signals}
\label{tab:txcompressao}
\begin{tabular}{l||cccccc}
\hlineB{2}
\multirow{2}{*}{\textbf{Signal}} & \multicolumn{3}{c}{\textbf{Num. of delivered packets}} &\hspace{-0.1cm}& \multicolumn{2}{c}{\textbf{Compression ratio} (\%)} \\ \cline{2-4}\cline{6-7}
& \textbf{
Original} & \textbf{LS} & \textbf{LL} & \hspace{-0.1cm}& \textbf{LS} & \textbf{LL} \\ \hlineB{2}
\textbf{ECGDB} & 8.863 & 5.589 & 5.757 &\hspace{-0.1cm}& 36.94 & 35.04 \\
\textbf{ECG} & 7.586 & 4.879 & 4.879 &\hspace{-0.1cm}& 36.68 & 36.68\\
\textbf{PPG} & 6.483 & 5.011 & 6.016 &\hspace{-0.1cm}& 22.70 & 7.20 \\ 
\textbf{Temp.} & 120 & 1 & 2 &\hspace{-0.1cm}& 99.16 & 97.50 \\\hlineB{2}    
\end{tabular}
\end{threeparttable}
\end{table}

GROWN got higher data compression ratio (PCR) mainly for highly correlated samples, as shows Table~\ref{tab:txcompressao}. We noted  that it greatly decreases the number of transmitted packets, and reduces the energy consumption. On the temperature signal, GROWN achieved a PCR of 99.16\% for LS, and 97.50\% for LL, as its  high correlated consecutive samples. Meanwhile, on
the ECG signal, it got a PCR of 35.68\% for both methods, as the correlation between consecutive samples exceeded the threshold of 1, and obliges the transmission of the same amount of information. As this behavior occurs less frequently with PPG signal, GROWN achieved the lowest PCR on it.

\section{Conclusion}
\label{sec:con}

\agn{This paper presented GROWN, a mechanism to save energy and increase the lifetime of wearable devices in wireless body networks. It manages real-time data compression in the sensing device by techniques employed in WSNs, so it allows devices to achieve an energy efficient operation. Experiments evaluated the GROWN effectiveness, and the results have demonstrated its ability to manage data compression and decompression in real-time in WBAN. GROWN has achieved an energy efficiency up to 53.73\% with a maximum
\agnc{delay}
of 55ms. Future works we will investigate GROWN availability in the face of data losses in the transmission, and its reliability with other wireless communication technologies. We also intend to compare GROWN with other mechanisms.}

\vspace{-0.1cm}
\section*{Acknowledgment}
\agnc{We would like to acknowledge the support of the Brazilian Agency CNPq in the Universal Project, grant \#436649/2018-7.}

\vspace{-0.1cm}
\bibliographystyle{IEEEtran}
\bibliography{latincom.bib}

\begin{thebibliography}{10}
\providecommand{\url}[1]{#1}
\csname url@samestyle\endcsname
\providecommand{\newblock}{\relax}
\providecommand{\bibinfo}[2]{#2}
\providecommand{\BIBentrySTDinterwordspacing}{\spaceskip=0pt\relax}
\providecommand{\BIBentryALTinterwordstretchfactor}{4}
\providecommand{\BIBentryALTinterwordspacing}{\spaceskip=\fontdimen2\font plus
\BIBentryALTinterwordstretchfactor\fontdimen3\font minus
  \fontdimen4\font\relax}
\providecommand{\BIBforeignlanguage}[2]{{%
\expandafter\ifx\csname l@#1\endcsname\relax
\typeout{** WARNING: IEEEtran.bst: No hyphenation pattern has been}%
\typeout{** loaded for the language `#1'. Using the pattern for}%
\typeout{** the default language instead.}%
\else
\language=\csname l@#1\endcsname
\fi
#2}}
\providecommand{\BIBdecl}{\relax}
\BIBdecl

\bibitem{tavares2020traffic}
T.~Tavares, M.~Nogueira, D.~Ros{\'a}rio, A.~Santos, and E.~Cerqueira,
  ``{Traffic Model Based on Autoregression for PPG Signals in Wearable
  Networks},'' \emph{IEEE Networking Letters}, vol.~2, no.~2, pp. 49--53, 2020.

\bibitem{movassaghi2014wireless}
S.~Movassaghi, M.~Abolhasan, J.~Lipman, D.~Smith, and A.~Jamalipour, ``Wireless
  body area networks: A survey,'' \emph{IEEE Communications Surveys \&
  Tutorials}, vol.~16, no.~3, pp. 1658--1686, 2014.

\bibitem{resque2019assessing}
P.~Resque, S.~Pinheiro, D.~Ros{\'a}rio, E.~Cerqueira, A.~Vergutz, M.~Nogueira,
  and A.~Santos, ``{Assessing Data Traffic Classification to Priority Access
  for Wireless Healthcare Application},'' in \emph{IEEE LATINCOM 2019}.\hskip
  1em plus 0.5em minus 0.4em\relax IEEE, 2019, pp. 1--6.

\bibitem{liao2018mutual}
Y.~Liao, M.~Leeson, Q.~Cai, Q.~Ai, and Q.~Liu, ``Mutual-information-based
  incremental relaying communications for wireless biomedical implant
  systems,'' \emph{Sensors}, vol.~18, no.~2, p. 515, 2018.

\bibitem{giorgi2017combined}
G.~Giorgi, ``A combined approach for real-time data compression in wireless
  body sensor networks,'' \emph{IEEE Sensors Journal}, vol.~17, no.~18, pp.
  6129--6135, 2017.

\bibitem{tsai2018efficient}
T.-H. Tsai and W.-T. Kuo, ``{An Efficient ECG Lossless Compression System for
  Embedded Platforms With Telemedicine Applications},'' \emph{IEEE Access},
  vol.~6, pp. 42\,207--42\,215, 2018.

\bibitem{marcelloni2010enabling}
F.~Marcelloni and M.~Vecchio, ``Enabling energy-efficient and lossy-aware data
  compression in wireless sensor networks by multi-objective evolutionary
  optimization,'' \emph{Information Sciences}, vol. 180, no.~10, pp.
  1924--1941, 2010.

\bibitem{azar2018performance}
J.~Azar, A.~Makhoul, R.~Darazi, J.~Demerjian, and R.~Couturier, ``On the
  performance of resource-aware compression techniques for vital signs data in
  wireless body sensor networks,'' in \emph{IEEE Middle East and North Africa
  Communications (MENACOMM)}.\hskip 1em plus 0.5em minus 0.4em\relax IEEE,
  2018, pp. 1--6.

\bibitem{antonopoulos2016resource}
C.~P. Antonopoulos and N.~S. Voros, ``Resource efficient data compression
  algorithms for demanding, wsn based biomedical applications,'' \emph{Journal
  of biomedical informatics}, vol.~59, pp. 1--14, 2016.

\bibitem{azar2018using}
J.~Azar, R.~Darazi, C.~Habib, A.~Makhoul, and J.~Demerjian, ``Using dwt lifting
  scheme for lossless data compression in wireless body sensor networks,'' in
  \emph{14th Int. Wireless Communications \& Mobile Computing Conference
  (IWCMC)}.\hskip 1em plus 0.5em minus 0.4em\relax IEEE, 2018, pp. 1465--1470.

\bibitem{deepu2017hybrid}
C.~J. Deepu, C.-H. Heng, and Y.~Lian, ``A hybrid data compression scheme for
  power reduction in wireless sensors for iot,'' \emph{IEEE trans. on
  biomedical circuits and systems}, vol.~11, no.~2, pp. 245--254, 2017.

\bibitem{schoellhammer2004lightweight}
T.~{Schoellhammer}, B.~{Greenstein}, E.~{Osterweil}, M.~{Wimbrow}, and
  D.~{Estrin}, ``Lightweight temporal compression of microclimate datasets
  [wireless sensor networks],'' \emph{29th Annual IEEE International Conference
  on Local Computer Networks}, pp. 516--524, Nov 2004.

\bibitem{sweldens1998lifting}
W.~Sweldens, ``The lifting scheme: A construction of second generation
  wavelets,'' \emph{SIAM journal on mathematical analysis}, vol.~29, no.~2, pp.
  511--546, 1998.

\bibitem{teuhola1978compression}
J.~Teuhola, ``A compression method for clustered bit-vectors,''
  \emph{Information processing letters}, vol.~7, no.~6, pp. 308--311, 1978.

\bibitem{marcelloni2008simple}
F.~Marcelloni and M.~Vecchio, ``A simple algorithm for data compression in
  wireless sensor networks,'' \emph{IEEE comm. letters}, vol.~12, no.~6, 2008.

\bibitem{mitdb2005}
G.~B. Moody and R.~G. Mark, ``{MIT-BIH Arrhythmia Database},''
  {https://physionet.org/content/mitdb/1.0.0/}, 2005, [Online]. Acessed in Jan.
  2020.

\bibitem{patel2013simulation}
D.~Patel, V.~Bhogan, and A.~Janson, ``Simulation and comparison of various
  lossless data compression techniques based on compression ratio and
  processing delay,'' \emph{IJCA}, vol.~81, no.~14, 2013.

\bibitem{zordan2014performance}
D.~Zordan, B.~Martinez, I.~Vilajosana, and M.~Rossi, ``On the performance of
  lossy compression schemes for energy constrained sensor networking,''
  \emph{ACM TOSN}, vol.~11, no.~1, pp. 1--34, 2014.

\bibitem{parks2007ohms}
J.~E. Parks, ``{Ohms Law III Resistors in Series and Parallel},''
  \emph{Department of Physics and Anatomy, University of Tennessee}, 2007.

\bibitem{gatouillat2018building}
A.~Gatouillat, B.~Massot, Y.~Badr, E.~Sejdi{\'c}, and C.~Gehin, ``Building
  iot-enabled wearable medical devices: an application to a wearable,
  multiparametric, cardiorespiratory sensor,'' 2018.

\bibitem{CC2540TI2020}
T.~Instruments, ``{CC2540 Bluetooth® Low Energy wireless MCU with USB},''
  \url{https://www.ti.com/product/CC2540}, 2013, (accessed March 1, 2020).

\end{thebibliography}

\end{document}